\documentclass[sigconf]{acmart}

\usepackage{tikz,graphicx,xcolor}
\usepackage{gensymb}
\usepackage{textcomp}
\usepackage{float,dblfloatfix}
\usepackage{verbatim,array,enumitem}
\usepackage{algorithm,algpseudocode,breqn,amsmath}
\usepackage{adjustbox,diagbox,threeparttable,tablefootnote,colortbl}
\usepackage{booktabs,tabularx,tabu,multirow,multicol,hhline}
\usepackage{caption,subcaption}
\PassOptionsToPackage{hyphens}{url} 
\usepackage{hyperref}
\usepackage[capitalise,noabbrev]{cleveref} 
\crefname{appsec}{Appendix}{Appendices} 
\usepackage[misc]{ifsym}
\usepackage[utf8]{inputenc}
\usepackage{booktabs}
\usepackage{lipsum}
\usepackage{tabularray}
\usepackage{titlesec}
\usepackage{longtable}
\usepackage{mdframed}
\usepackage{tcolorbox}
\usepackage{multicol}
\usepackage{framed}

\usepackage{listings}
\usepackage{xcolor}

\lstdefinestyle{python}{
    language=Python,
    basicstyle=\ttfamily\footnotesize,
    keywordstyle=\color{blue},
    stringstyle=\color{green!40!black},
    commentstyle=\color{gray},
    showstringspaces=false,
    breaklines=true,
    frame=single,
    captionpos=b,
    numbers=left,
    numberstyle=\tiny\color{gray},
    stepnumber=1,
    numbersep=5pt,
}

\copyrightyear{2025}
\acmYear{2025}
\acmConference[CODASPY '25]{Proceedings of the ACM Conference on Data and Application Security and Privacy 2025}{June 4--June 6, 2025}{Pittsburgh, PA, USA}
\acmBooktitle{Proceedings of the ACM Conference on Data and Application Security and Privacy 2025 (CODASPY '25), June 4--June 6, 2025, Pittsburgh, PA, USA}

\titleformat{\subsubsection}
  {\normalfont\normalsize\itshape}{\thesubsubsection}{1em}{}

\titlespacing*{\subsubsection}{0pt}{1.25ex plus 1ex minus .2ex}{0.75ex plus .2ex}

\begin{document}

\title[Why You've Got Mail]{Why You've Got Mail: Evaluating Inbox Privacy Implications of Email Marketing Practices in Online Apps and Services}

\author{Scott Seidenberger}
\affiliation{%
  \institution{University of Oklahoma}
  \city{Norman}
  \state{OK}
  \country{USA}
}
\email{seidenberger@ou.edu}

\author{Oluwasijibomi Ajisegiri}
\affiliation{%
  \institution{University of Oklahoma}
  \city{Norman}
  \state{OK}
  \country{USA}
}
\email{oluwasijibomi.ajisegiri@ou.edu}

\author{Noah Pursell}
\affiliation{%
  \institution{University of Oklahoma}
  \city{Norman}
  \state{OK}
  \country{USA}
}
\email{noah.a.pursell-1@ou.edu}

\author{Fazil Raja}
\affiliation{%
  \institution{University of Oklahoma}
  \city{Norman}
  \state{OK}
  \country{USA}
}
\email{fazilraja13@gmail.com}

\author{Anindya Maiti}
\affiliation{%
  \institution{University of Oklahoma}
  \city{Norman}
  \state{OK}
  \country{USA}
}
\email{am@ou.edu}

\begin{abstract}

This study explores the widespread perception that personal data, such as email addresses, may be shared or sold without informed user consent, investigating whether these concerns are reflected in actual practices of popular online services and apps. Over the course of a year, we collected and analyzed the source, volume, frequency, and content of emails received by users after signing up for the 150 most popular online services and apps across various sectors. By examining patterns in email communications, we aim to identify consistent strategies used across industries, including potential signs of third-party data sharing. This analysis provides a critical evaluation of how email marketing tactics may intersect with data-sharing practices, with important implications for consumer privacy and regulatory oversight. Our study findings, conducted post-CCPA and GDPR, indicate that while no unknown third-party spam email was detected, internal and authorized third-party email marketing practices were pervasive, with companies frequently sending promotional and CRM emails despite opt-out preferences. The framework established in this work is designed to be scalable, allowing for continuous monitoring, and can be extended to include a more diverse set of apps and services for broader analysis, ultimately contributing to transparency in email address privacy practices.

\end{abstract}

\keywords{Email Privacy, Email Marketing, Unsolicited Email, Spam.}

\maketitle

\section*{Taxonomies}
\label{sec-tax}

We distinguish email sources based the following classifications:

\begin{itemize}[leftmargin=*]
    \item \textbf{Internal:} Emails sent directly by the service's own infrastructure. They typically should pass Sender Policy Framework (SPF) and DomainKeys Identified Mail (DKIM) checks.
    \item \textbf{Authorized Third-Party (ATP):} Emails sent on behalf of the service by authorized providers, authenticated by the service. They typically should also pass SPF and DKIM checks. 
    \item \textbf{Unknown Third-Party (UTP):} Emails from unauthorized or unknown entities. If such an email is masqueraded as coming from the service, it typically will fail SPF and DKIM checks.
\end{itemize}

\noindent
Unsolicited emails are grouped into these broad categories:

\begin{itemize}[leftmargin=*]
    \item \textbf{Service-Originated Spam (SOS):} Unsolicited internal and ATP emails that pass SPF and DKIM checks.
    \item \textbf{Unknown or Unauthenticated Sender Spam (UUSS):} Emails from UTP sources and emails that fail SPF \emph{and} DKIM checks (even if the header indicates service-originated in the \texttt{From:} field).
\end{itemize}

\noindent
Emails are further categorized by purpose:

\begin{itemize}[leftmargin=*]
    \item \textbf{CRM:} Customer Relationship Management emails aimed at engagement and relationship-building.
    \item \textbf{Promotional:} Emails with overt selling intent, such as discounts or offers.
    \item \textbf{Alerts:} Informational emails, such as account notifications or updates.
\end{itemize}

\section{Introduction}
\label{sec-intro}

Public perceptions of data privacy, particularly concerning personal data such as email addresses, can be significantly shaped by high-profile data breaches, feelings of powerlessness, and general misunderstandings. Isolated incidents of major data breaches contribute to the growing concern among users about the security of their personal information. For instance, the Yahoo data breach in 2013–2014 affected all 3 billion user accounts, exposing email addresses and other personal data~\cite{yahoohack}. Similarly, the Equifax breach in 2017 compromised the personal information of approximately 147 million people, including email addresses and Social Security numbers~\cite{equifaxhack}. These incidents have heightened public awareness of vulnerabilities in corporate data handling practices.
Such breaches exacerbate the concept of ``digital resignation'' where individuals feel powerless to control their personal information, accepting that their email addresses and other data may be shared or exposed without their consent~\cite{draper2019corporate}. This resignation is further fueled by past corporate practices that shared data with third-parties~\cite{englehardt2018never}, before data privacy laws such as GDPR~\cite{voigt2017eu} and CCPA~\cite{pardau2018california} were enacted. 
Hoofnagle et al.~\cite{hoofnagle2012privacy} found that users are often unaware of the extent to which their email addresses are collected and shared with third-parties, increasing feelings of vulnerability and mistrust.
Furthermore, %
Acquisti et al.~\cite{acquisti2015privacy} found that even users who express privacy concerns can themselves engage in behaviors that compromise their email data, often due to a lack of viable alternatives or understanding.

\emph{A significant challenge for users is the difficulty of identifying which app or service may have leaked or shared their email address when they start receiving unsolicited spam or phishing emails.} This issue arises because most users generally reuse the same email address across multiple online services and apps. Consequently, when unsolicited communications are received from third-parties, users may not be able to determine the specific source responsible for any unauthorized sharing or leakage of their email information. %
This difficulty in identifying the source of leaked or shared email addresses is further exacerbated by pervasive email marketing practices, which were often connected to the sharing of email and other personal data with third-parties. %
Companies often engaged in email marketing strategies that involve partnering with affiliates, utilizing data brokers, or participating in advertising networks to broaden their outreach~\cite{sherman2021data}. These practices frequently entailed sharing or selling users' email addresses, sometimes without informed consent or clear disclosure. As a result, users could receive unsolicited promotional emails from entities with whom they have no direct relationship, obscuring the boundaries of accountability. The interconnection between email marketing and email sharing practices not only increases the burden of unsolicited communications~\cite{vacek2014email,agema2015death} but also undermines users' ability to trust that their personal information is handled responsibly.

After the introduction of data privacy laws like GDPR and CCPA, there still remains a significant gap towards systematically auditing online services and apps for their email address sharing and email marketing practices. Email marketing strategies have become increasingly sophisticated, often blurring the lines between legitimate marketing and unauthorized data sharing with third-parties. Without robust auditing mechanisms, it's challenging to assess compliance with these regulations, hold companies accountable for improper handling of email data, or impose consequences for sending unsolicited emails despite user opt-outs. Developing such a framework is crucial for enhancing transparency and accountability in email marketing practices, ultimately improving user perceptions by rebuilding trust in how their personal information is managed.

Towards addressing this gap, %
our study offers a structured and scalable framework to assess how online services and apps handle users' email addresses, from the perspective of an user inbox. Our contributions include the following key steps and insights:
\begin{itemize}[leftmargin=*]
    \item \textbf{Audit of Online Services and Apps}: We conducted a year-long audit of email communications received after registering with the top 150 popular online services and apps, opting out of any optional communication during the registration process. No unknown or unauthenticated sender spam (UUSS) emails were detected within this time frame, suggesting that opting out during account creation is very effective in limiting unsolicited emails from unknown and unauthenticated entities.
    
    \item \textbf{Email Marketing Analysis}: Despite no UUSS emails, we collected 4,847 emails received from these online services and apps, directly (internal) or from authorized third-parties (ATP). Our analysis revealed that companies still frequently used customer relationship management (CRM), alert, and promotional emails to engage users, even when opting out of promotional communications. This highlights the persistence of well-established marketing strategies, majority of which may be classified as service-originated spam (SOS).
    
    \item \textbf{Cross-Sector Insights}: We identified significant differences in email marketing practices across sectors. Consumer industries, such as omnichannel retail, heavily relied on promotional campaigns, while financial services prioritized CRM and alert emails. Some of these sector-specific behaviors offer insights into varying risks to inbox privacy. %
\end{itemize}

These findings contribute to a broader understanding of email marketing and data sharing practices, promoting transparency and accountability in how online services and apps handle users' email addresses.

\begin{figure*}[t]  %
    \centering
    \includegraphics[width=0.86\textwidth]{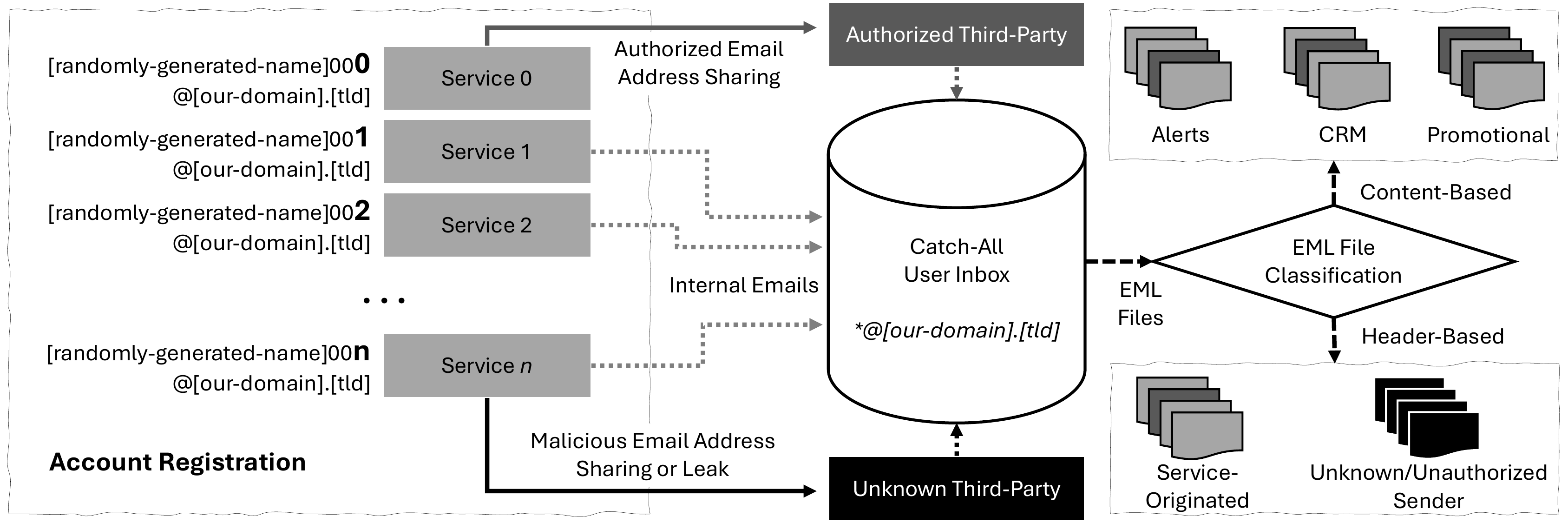}
    \caption{Framework for email collection, processing, and classification to analyze inbox privacy across services.}
    \label{fig:framework} %
\end{figure*}

\section{Related Work}
\label{sec-related}

Email and inbox privacy has become a critical issue in the wake of increased data breaches and unauthorized sharing of personal information. Englehardt et al.~\cite{englehardt2018never} demonstrated the widespread use of email tracking, allowing third parties to gather user behavior data without consent. Their findings emphasize how email addresses, once shared, become tools for covert tracking, often leading to privacy violations. Similarly, Hoofnagle et al.~\cite{hoofnagle2012privacy} found that most users are unaware of how their email data is shared with third parties for advertising purposes, heightening concerns about privacy infringement. %
Plice et al. ~\cite{plice2008spam} provided an economic analysis of unwanted commercial messages and their dependence on weak privacy controls. These studies highlight the necessity of open auditing mechanisms to protect email privacy.

Email marketing remains a central tool for customer engagement, but its effectiveness is tied to timing, frequency, and personalization. Thomas et al.~\cite{doi:10.1177/10949968221095552} found that emails with lower levels of overt persuasion, such as CRM emails, tend to have higher open and click-through rates compared to promotional emails, aligning with the \textit{Persuasion Knowledge Model}~\cite{friestad1994persuasion}. Similarly, Nuutinen~\cite{nuutinen2015increase} emphasized the role of segmentation and personalization in improving email performance, noting that content relevance significantly impacts open and conversion rates.
The importance of timing has also been explored extensively. Araujo et al.~\cite{araujo2022novel} proposed a machine learning approach using regression models like Random Forest and LSTM to predict the optimal send times for marketing emails. Their study demonstrated that predictive algorithms could improve open rates by sending emails when users are most likely to engage, highlighting the importance of behavioral analytics in email marketing. Abrahams et al.~\cite{abrahams2010multi}, in their multi-industry study of large U.S. franchises, found that industries like hospitality and food service were prolific in email frequency, often sending 2-4 emails per week, compared to sectors like automotive, which sent emails infrequently. These findings provide valuable benchmarks for industry-specific marketing practices.
Further research by Yang et al.~\cite{yang2019post} examined the impact of post-stay email marketing in the hotel industry, identifying how personalized, interactive emails increased customer retention and revisit intention. This aligns with the findings from~\cite{budac2016theoretical}, which highlighted the role of A/B testing, personalization, and frequency in determining email effectiveness across different industries.

Existing literature highlights the need for a balance between effective email marketing and safeguarding inbox privacy. While predictive models and personalized strategies can optimize engagement, they must be deployed ethically to avoid violating user privacy. Prior works emphasize that compliance with data protection laws, along with the adoption of transparent email marketing practices, is critical for maintaining consumer trust. Our study builds on this foundation by analyzing the email marketing and data-sharing practices of popular apps and services. Unlike prior works, which often did in-depth analysis of emails pertaining to a single company or industry type, our study instead focuses on a breadth of services and apps across various sectors.

\section{Data Collection Framework}
\label{sec-goals}

Our framework (\cref{fig:framework}) is based on the assumption that if email addresses are leaked, shared, or sold by the services or apps we signed up for, those addresses would eventually start receiving spam or unsolicited emails from third-parties. This assumption allows us to passively detect unauthorized sharing of user data through the presence of unsolicited communications, which are typically linked to marketing activities that involve purchasing user data from third-party vendors. It is important to note, however, that this approach has its limitations. Specifically, we cannot detect instances where email addresses are leaked, shared or sold to third-parties if they do not subsequently send spam or other detectable communications.

To comprehensively evaluate the email sharing and marketing practices across different sectors, we selected a diverse set of widely used online services and mobile applications, as follows:

\begin{itemize}%
    \item \textbf{Top 100 Online Services:}
   We compiled a list of the most popular websites worldwide based on sources such as Cisco Umbrella\footnote{\url{https://umbrella-static.s3-us-west-1.amazonaws.com/index.html}} and Statista\footnote{\url{https://www.statista.com/}}. Services that were consistently ranked across the two sources were prioritized for inclusion in the final list. %

   \item \textbf {Top 50 Mobile Applications:}
   Similarly, a list of the top 50 mobile applications was compiled by analyzing app rankings from Google Play Store and Apple App Store. Notably, mobile apps that were already included in the top 100 online services (e.g., Facebook) were excluded to ensure non-redundant data. This ensures that our dataset captures a broader spectrum of both online services and apps across platforms. %

\end{itemize}

\textbf{Account Registration.}
After finalizing the list of online services and apps, we began the account registration process for each, utilizing standardized contact information. 
For each identified entity (online service and mobile app), a unique account was created following a standardized registration procedure. Unique email addresses were generated in the following format: \textbf{[randomly-generated-name]$xxx$@[our-domain].[tld]}, with specific ranges of $xxx$ assigned as follows:
\begin{itemize}[leftmargin=*]
        \item 000-099 for the 100 online services.
        \item 100-149 for the 50 mobile apps.
\end{itemize}

This design of our framework facilitates clear differentiation between service types and ensures that each service and app has a unique identifier throughout the experiment. During registration, no additional subscriptions, email notifications, or promotional materials were selected. This was essential for maintaining consistency in data collection, allowing us to focus on analyzing unsolicited marketing emails from the service and app providers.

All received emails were collected in a single catch-all inbox, making it easier to manage, store, and scale the data collection. During the entire experiment, we did not click on any unsubscribe links in the received emails, which also helped maintain a consistent approach to data collection without influencing the outcome.
At the end of the 361-day data collection period, all received emails were saved as EML files, containing the email headers, subject, body and any additional metadata, for our analyses.

\textbf{Email Content Classification.} To classify the content of the emails, we rely on definitions provided by Thomas et al. \cite{doi:10.1177/10949968221095552}, where emails were classified as either promotional, CRM, or alerts. These email types convey distinct messages with differing levels of persuasion. It is the appropriate combination of these different types of messages at the right time that drive click-through and eventual conversion, or conversely, abandonment, by the potential customer. To classify the emails that we collected, we used an in-context classification approach with a general purpose LLM (Llama3.1-8b-instruct \cite{llama3modelcard}) with structured output. This approach has been shown to achieve human-level performance on sufficiently large LLMs with clear and concise rules~\cite{brown_language_2020}. The full procedure is available in~\cref{appendix-llm-classification}.  There were 3,346 (69.0\%) promotional emails, 1,393 (28.7\%) CRM emails, 92 (1.89\%) alert emails, and 16 (0.33\%) that were unable to be classified as they were unparseable. 

The use of LLMs as a rater for content classification tasks such as this is an area of active research \cite{theelen2024doing, chae2023large}, and recently has been successfully applied to spam email classification tasks \cite{rojas2024zero}. For this task, however, which relies on the LLM to perform comparably to a human rater, we enlisted 5 independent researchers from our institution to classify 50 randomly selected emails using the same instructions given to the LLM. Since there is no ground-truth, we instead evaluate whether the LLM-human inter-rater reliability (IRR) falls within the range of human-human IRR. The average Cohen’s Kappa among human-human pairs is $\kappa=0.406$ (moderate agreement), and the average LLM-human Kappa is $\kappa=0.479$ (moderate agreement). In other words, the LLM achieves a level of agreement with humans that not only exceeds the human-human average, but also places it well within the variability observed among human raters themselves. This suggests that the LLM is performing comparably to human raters, demonstrating its viability as a reasonable evaluator in this classification task.

Our framework and following analyses address key aspects of data privacy laws, including the right to opt-out of marketing communications and restrictions on unauthorized data sharing practices, and can be extended to evaluate compliance across more diverse regulatory environments.

\section{Findings}
\label{sec-results}

\subsection{Descriptive Statistics}

Over the 361-day collection period, we received 4,847 emails from 109 of the 150 services and apps, post registration (\cref{table:all-domains} in \cref{appendix-domains}). Of those emails received, 99.96\% passed SPF checks and 81.64\%	passed DKIM checks. All emails that passed DKIM checks also passed SPF checks. Sender Policy Framework (SPF) verifies the sender’s IP address to ensure it is authorized by the domain owner, while DomainKeys Identified Mail (DKIM) adds a cryptographic signature to emails, verifying that the message hasn't been altered in transit and is indeed from the claimed sender. 
Although ideally, both SPF and DKIM should pass, the successful verification of either check indicates that the email was authorized by the service with high confidence. A detailed manual review of the three emails that failed both SPF and DKIM revealed that these failures were likely due to DNS or server misconfigurations rather than malicious intent.

\begin{leftbar} \noindent
Post inspection of the emails received by their root domains and the email security checks via SPF and DKIM, we conclude that all emails received were from the original companies' domains with which we had made accounts and no UUSS emails were detected in this dataset.
\end{leftbar}

\begin{figure}[b]
    \centering
    \includegraphics[width=\linewidth]{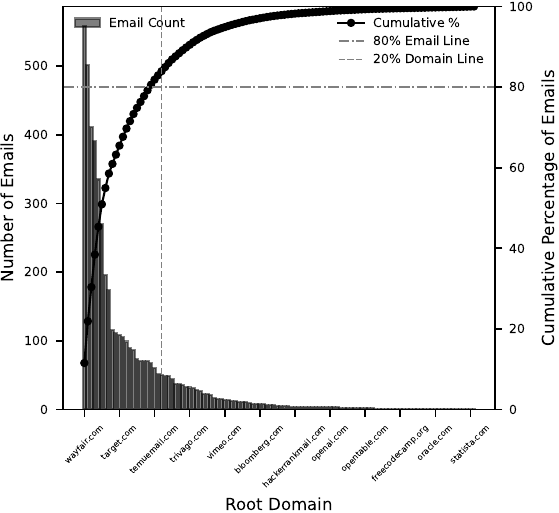}
    \caption{Pareto diagram of total email volume.}
    \label{fig:volume-pareto} %
\end{figure}

However, further analysis reveals that the origin of the emails tell a more complex story of the lineage and custody of user data. 
Initial descriptive statistics indicate that there is not a well-defined parametric distribution that determines the volume of emails received. The mean was 43 emails per domain with a median of 5, a skewness of 3.55 showing the heavy right tail, and kurtosis of 12.92. The data's high skewness and kurtosis suggest a complex, highly skewed, and heavy-tailed nature that is not well captured by parametric distributions. \cref{fig:volume-pareto} shows that a few large senders make up the dominant amount of email volume received, with the cumulative distribution following the 80-20 Pareto principle, where the total volume of emails received are dominated by a few senders. Just the top 10 root domains make up 63.23\% of the total volume.

The emails received do not just vary in the volume sent by companies, but also in the content of the messages being sent. Previous literature shows that strategic email marketing campaigns differ in both \textit{what} the sender says (the content) and \textit{when} they say it (temporal dimension) \cite{thomas2022email}. Additionally, previous work has shown that not all companies engage equally in strategic email marketing \cite{abrahams2010multi}, with some being more sophisticated actors than others. Therefore, the following subsections of the analysis are structured as such: (1) We explore a unique contribution of our dataset to the literature on this topic, which is the sender's IP address and associated information about that sender through its Autonomous System Number (ASN). (2) A temporal analysis to show at what times and with what frequency emails are sent. (3) Clustering of companies based on their emails content and temporal patterns to determine where companies fall on their level of sophistication or general tactics. 

\begin{figure}[t]
    \centering
    \includegraphics[width=0.99\linewidth]{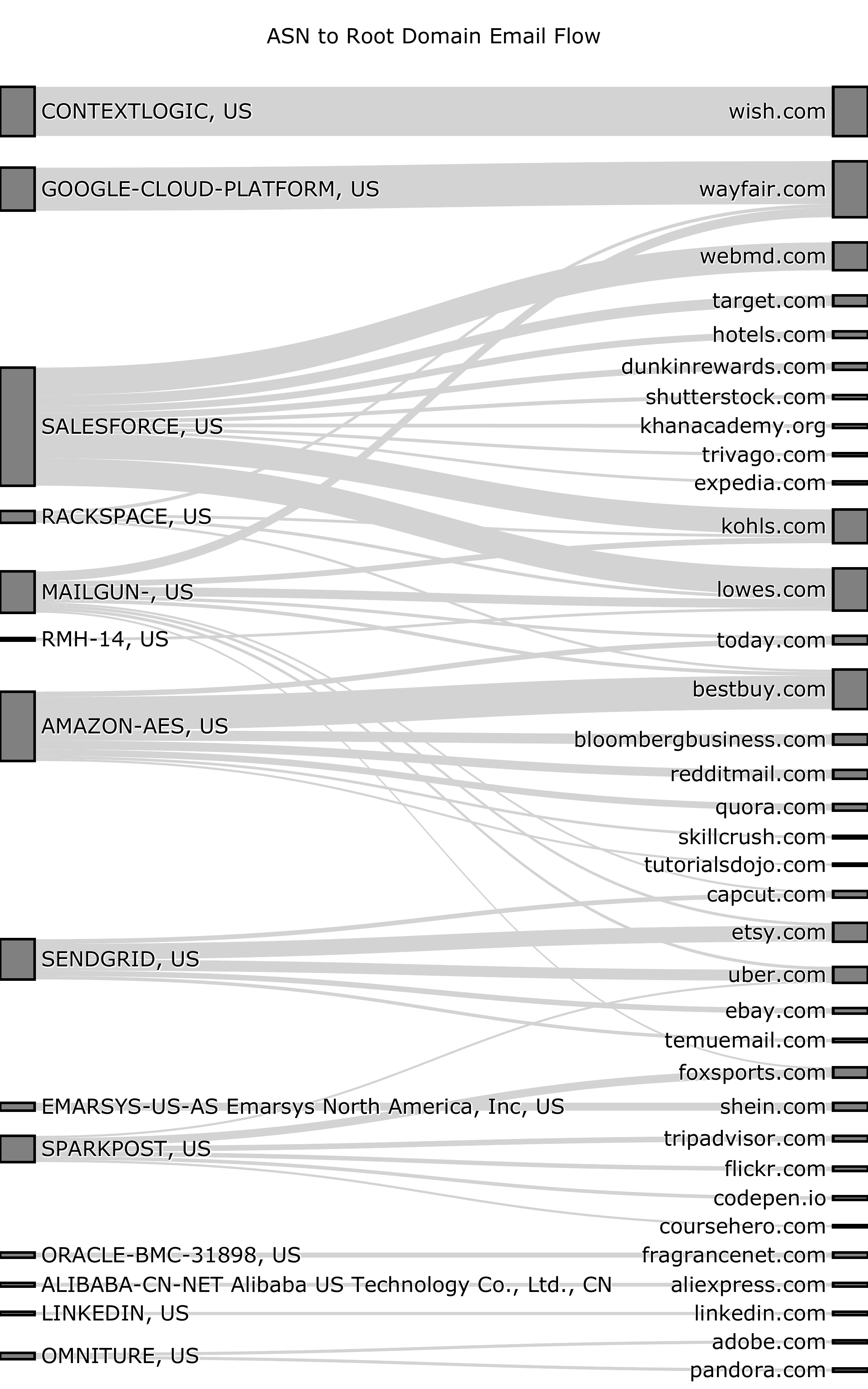}
    \caption{Sankey diagram of the top 50 flows.}
    \label{fig:sankey_diagram} %
\end{figure}

\subsection{Senders by Autonomous System}

We captured the IP address of the sender of each email, with that we then ran a lookup of the registered autonomous system for each of those IP addresses and were able to tie it back to the company that sent it. Because of this, we are able to identify that \textbf{most of the emails received come from authorized third-parties} that specializes in business-to-business marketing and CRM services like \textit{Salesforce} and \textit{Mailgun}. Like the overall email volume, the email volume by ASN also follows the Pareto principle, where a few large providers are responsible for the sending of the large majority of emails. Specifically, the top twenty percent, 8 ASNs, produced 89.35\% of the total email volume. \cref{fig:sankey_diagram} shows the evidence that most of the email marketing flow originates from a few large, third-party providers. 

However, understanding which companies use which email marketing service, or whether they use a third-party service at all, is not straightforward. Some companies send emails originating from their own registered ASN such as \textit{wish.com} (Wish is the ``doing-business-as'' name for ContextLogic), Alibaba Group, LinkedIn, and Adobe (Adobe acquired Omniture as part of their web analytics platform). Others employ specialized third-parties, chiefly \textit{Salesforce}, \textit{Mailgun}, \textit{Sendgrid}, and \textit{Sparkpost}. A third group had emails originate from cloud service providers like \textit{AWS}, \textit{GCP}, and \textit{Rackspace}. For those that originate from cloud service providers, using only the IP addresses we cannot ascertain what entity is managing the email campaigns at the application layer. 

\textbf{Sender Reputation.} While the domain-level analysis confirms authenticity and shows no evidence of unauthorized or unknown senders, the infrastructure-level analysis reveals a potentially more ambiguous trust environment. To deepen our understanding of the quality and trustworthiness of these sending infrastructures, we conducted an IP reputation analysis. We cross-referenced each unique email origination IP against a third-party IP reputation database~\cite{abuseIPDB}, which aggregates historical spam reports and other abuse metrics. In total, there were 612 spam reports associated with 84 unique IP addresses belonging to 11 ASNs, corresponding to 29 unique companies that were in our study. This allows us to assess the prevalence of reported spam activity associated with the IP ranges, from which we received emails, used by various ASNs.

\begin{figure}[b]
    \centering
    \includegraphics[width=\linewidth]{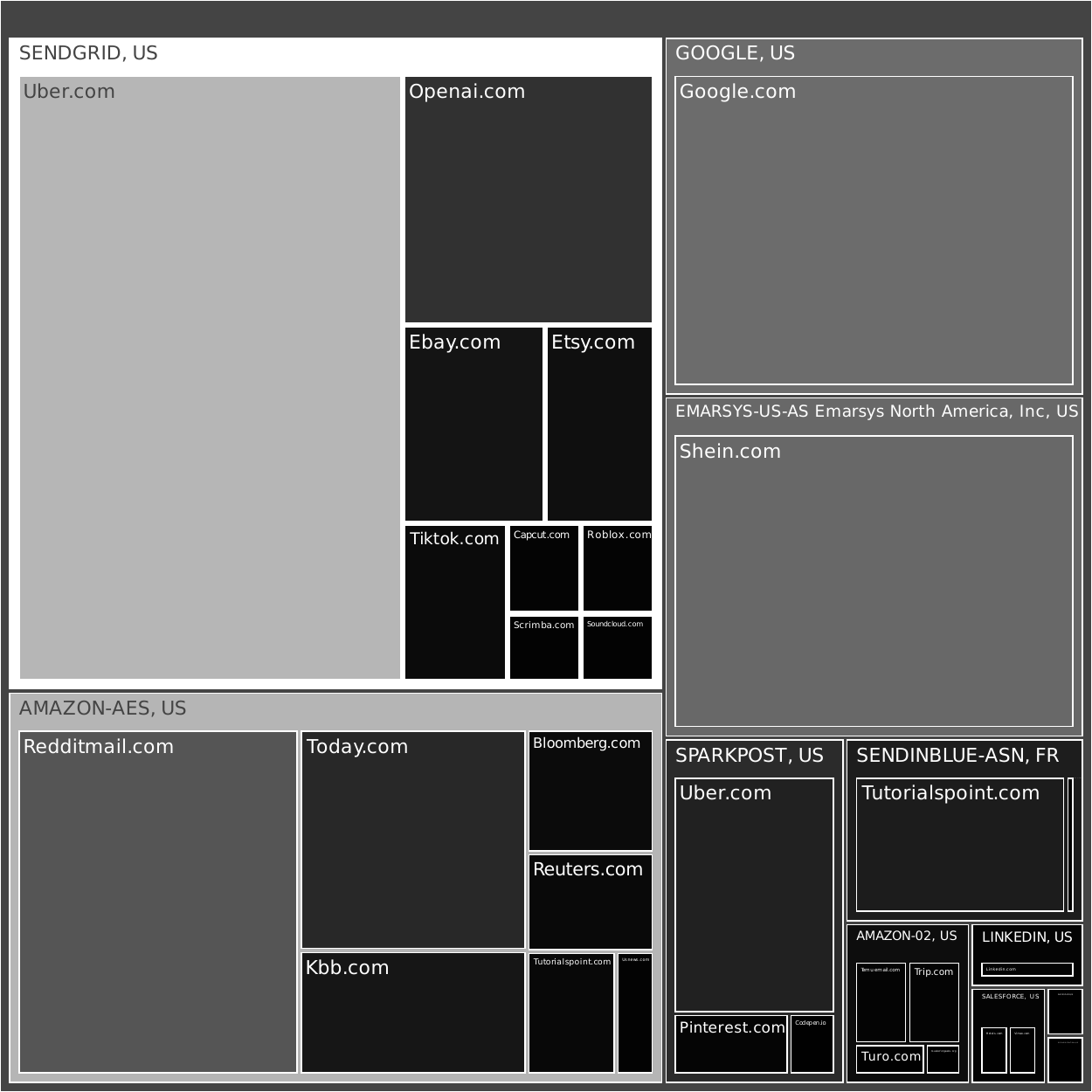}
    \caption{Hierarchical treemap of community reported spam on sender IPs, organized by ASN and associated company domains. Lighter shades indicate higher counts of spam reports, and vice versa.}
    \label{fig:sender-reputation}%
\end{figure}

By coupling the ASN-level volume analysis with the historical IP reputation data, we found that even well-known brands may rely, potentially unknowingly, on email infrastructure that carries varying levels of reported spam activity. The treemap visualization in \cref{fig:sender-reputation} arranges ASNs hierarchically, with each company's sending domain nested beneath the ASN that serviced it. The area of each box corresponds to the total number of spam reports for the associated IPs, while the grayscale intensity indicates the magnitude of these reports.

This hierarchical perspective reveals patterns and raises important questions for both research and policy. For instance, some large ASNs have a relatively high volume of reported spam across multiple well-known clients. In such cases, the individual brand may not be involved in illicit or spammy behaviors, yet they inherit the reputation challenges posed by their chosen infrastructure providers. Or, the community-writ-large will deem certain messages as spam and report them as abuse. These findings highlight that brand trust does not necessarily translate into a pristine underlying email infrastructure. Rather, the operational practices, vetting procedures, and network hygiene of intermediary email service providers matter a great deal.

\begin{leftbar} \noindent
Ultimately, these findings show that the email ecosystem is more complex and federated than it may appear on the surface. While the messages themselves may appear genuine and from the expected sender, the underlying infrastructure often involves multiple intermediaries and networks of varying reputations. 
\end{leftbar}

For end-users, this complexity raises concerns about data stewardship and privacy. When personal information and communication preferences flow through third-party service providers, some of which may have been associated with spam-related activity, users face heightened uncertainty about how their data is handled, who can access it, and to what end. This underscores the need for greater transparency and more privacy-centric policies from both the companies authorizing these messages and the infrastructure providers enabling their delivery.

\textbf{IP Hopping and Spam Reports.} To investigate whether companies associated with higher spam reports are more likely to use a larger number of IPs, a behavior that could suggest ``IP hopping'' to evade reputation-based spam filters, we examined the correlation between the number of unique IPs per company and the total spam reports associated with those IPs.

Our analysis revealed a strong positive correlation between the number of IPs used by a company and their associated spam reports. Pearson's correlation coefficient ($r=0.71, p<.0001$) indicates a linear relationship, where companies using more IPs tend to have higher volumes of spam reports. Spearman's rank correlation ($r_s=0.55, p=.002$) further supports this finding, suggesting a monotonic relationship even in the presence of potential non-linearity or outliers.

\begin{leftbar} \noindent
This result highlights that companies linked to higher spam reports may be employing a strategy of cycling through multiple IP addresses to distribute the reputational impact across their infrastructure. Such IP hopping practices can dilute the effects of blacklistings or abuse reports on any single IP address \cite{ramachandran2007filtering}, allowing these companies to maintain email delivery volumes despite the reputational damage incurred by individual IPs.
\end{leftbar}

\subsection{Temporal Analysis}

\begin{figure}[t]
    \centering
    \includegraphics[width=\linewidth]{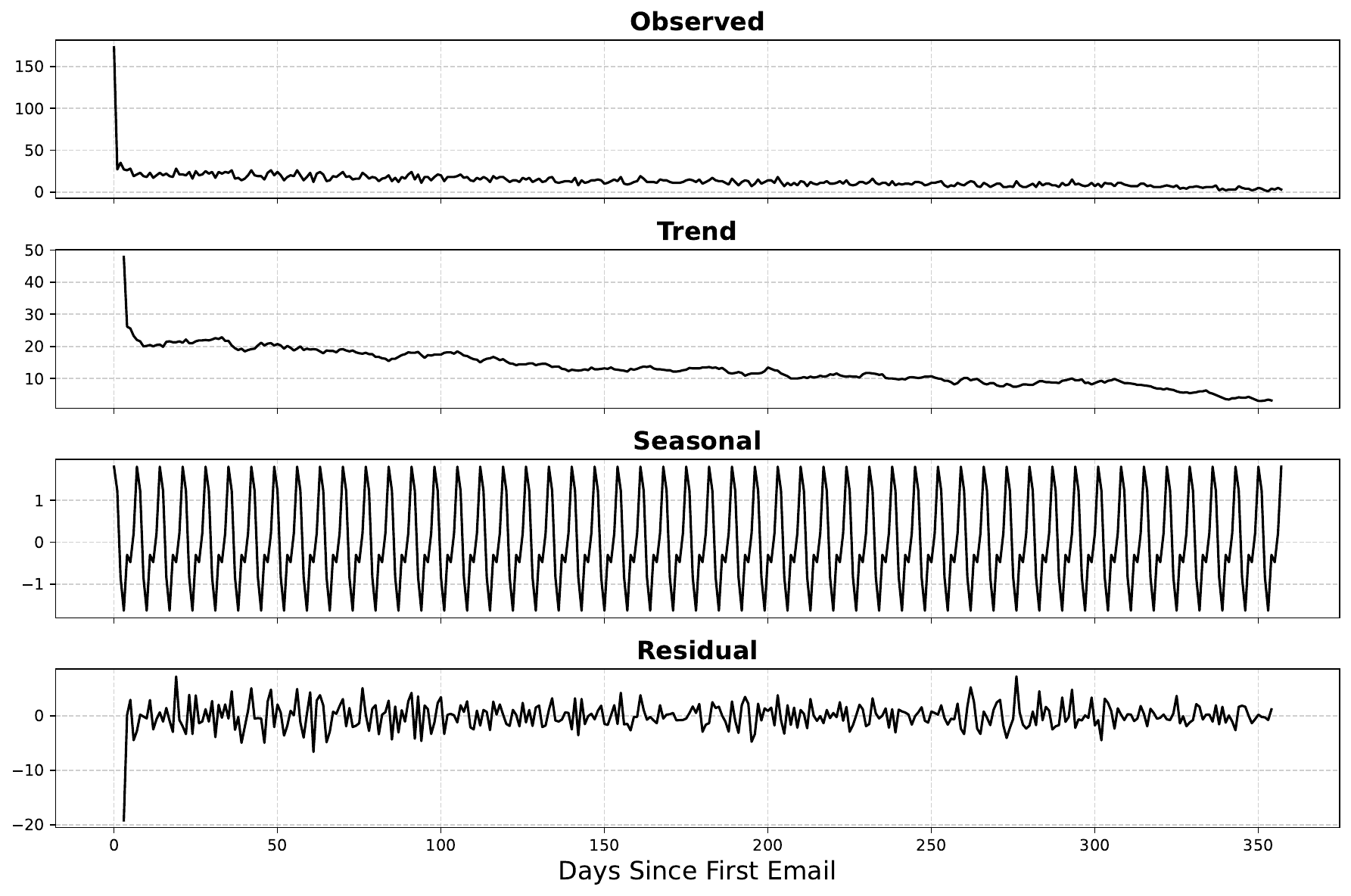}
    \caption{Additive time series decomposition of aggregate emails received.}
    \label{fig:additive-time-series} %
\end{figure}

The temporal dimension of this analysis examines both the timing (time of day and day of week) and the frequency of emails sent. First, we treat the incoming messages as a time series, decomposing it to understand the general trend and identify hypothesized seasonality as shown in \cref{fig:additive-time-series}. Since account creation dates differ across services and apps, we standardize the series by the number of days since the first email received, which coincides with the account creation date for each app. This standardization provides a uniform time scale for the analysis.

Applying a Fourier transform to the combined time series revealed dominant frequencies, identified at the $2\sigma$ level, corresponding to various periodic cycles: annual, semi-annual, quarterly, and weekly. The identified frequencies ($f$), magnitudes ($|X(f)|$), and periods ($T$) are as follows:

\begin{itemize}
    \item Annual cycle: $f = 0.0028$, $|X(f)| = 834.38$, $T \approx 362$ days
    \item Semi-annual cycle: $f = 0.0055$, $|X(f)| = 479.22$, $T \approx 181$ days
    \item Quarterly cycle: $f = 0.0138$, $|X(f)| = 308.50$, $T \approx 72.4$ days
    \item Weekly cycle: $f = 0.1436$, $|X(f)| = 393.39$, $T \approx 6.96$ days
\end{itemize}

\begin{figure*}[htbp]  %
    \centering
    \includegraphics[width=\textwidth]{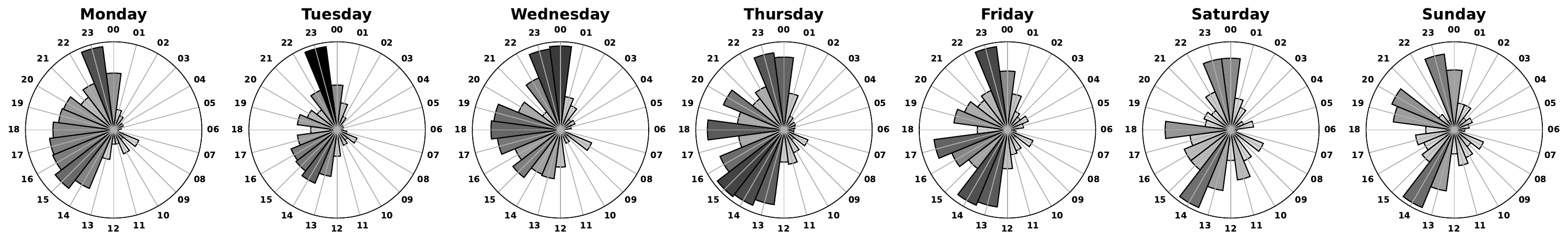}
    \caption{Email frequency by hour of day for each day of the week.}
    \label{fig:email_frequency} %
\end{figure*}

The most relevant frequency for our analysis is the weekly cycle ($\approx 7$ days), denoted by $f = 0.1436$ with a magnitude of $|X(f)| = 393.39$. This weekly periodicity corresponds closely with the expected operational and behavioral cycles observed in email-sending patterns, reflecting regular weekly activity. We use the 7-day period as the seasonality figure for the time series decomposition. 
\begin{leftbar} \noindent
The decomposition highlights a linear, decreasing trend in the number of emails received over time. Additionally, it shows that the weekly seasonal component explains only 1.17\% of the variance in the number of emails received, which is evidence of varied strategies being employed in when emails are sent. 
\end{leftbar}
The amount of seasonality in email sending behavior is mediated at the company level and will be explored in greater detail in following sections on the cross-sector and company-level analysis. 
Given that the observed weekly seasonality in sending patterns we look to specify the general time of day and day of week that email messages are sent. The polar heat maps shown in~\cref{fig:email_frequency} show that the most popular times messages are sent occur during the mid-week, Tuesday through Thursday, and emails are predominantly sent around midnight or during the afternoon work hours between 1300 and 1500 hours, specifically targeting when users are most active online. 

\subsection{Clustering}

The next task is to determine which companies exhibit similar email sending patterns, and then map those patterns to levels of sophistication from prior literature. To accomplish this unsupervised learning task, we first perform a principal component analysis (PCA) on all the features created from the previous analysis. Then, we perform K-Means to assign each company to a cluster. 

The analysis began by loading the relevant email and ASN data and selecting the temporal and behavioral features.

\begin{itemize}
    \item \textbf{Hourly Email Frequency}: We computed the count of emails sent by each company per hour of the day (0-23), aggregated across all days.
    \item \textbf{Weekly Seasonality}: We analyzed the weekly email sending pattern by determining the frequency of emails sent on each day of the week (0 = Monday, 6 = Sunday).
    \item \textbf{Professional Marketing Indicator}: Each ASN description was checked against known marketing services (e.g., Salesforce, Mailgun, SendGrid, SparkPost) to create a binary feature indicating the presence of professional marketing practices.
    \item \textbf{Email Content Classification}: We captured the distribution of email content (e.g., promotional, CRM, alert) for each company, allowing us to profile their communication style.
    \item \textbf{Total Email Volume}: The overall number of emails sent by each company was computed to gauge their communication scale.
\end{itemize}

These features were combined into a comprehensive feature set, standardized using \texttt{StandardScaler}, and subsequently reduced to five principal components using PCA, retaining over 80\% of the variance in the data. %

K-Means clustering was employed to group companies based on their PCA-transformed features. To determine the optimal number of clusters, we evaluated cluster validity using the Silhouette Score, a metric that measures how similar a point is to its assigned cluster compared to other clusters. We tested a range of cluster numbers (2-10) and identified %
that separating companies into two clusters yielded the highest silhouette score (0.831). %
Using the PCA loadings, we interpret the characteristics of each cluster to better understand the distinct email marketing behaviors exhibited by companies within each group. The average scores of each cluster on the principal components highlight the dominant features that distinguish one cluster from another.

\begin{itemize}%
    \item \textbf{Cluster 0: Low-Activity Generalists} \\ Cluster 0 exhibits negative scores on Component 1 and slight negative or near-zero scores on other components. This cluster represents companies with generally low email activity, characterized by a lack of pronounced patterns in sending frequency, content, or professional marketing service usage. The absence of significant influence from any principal component suggests that these companies do not heavily invest in sophisticated or high-volume email marketing strategies. Their approach may be minimalistic or highly targeted with low engagement, reflecting a more traditional or less dynamic email marketing approach.

    \item \textbf{Cluster 1: High-Volume Promotional Marketers} \\ Cluster 1 stands out with a highly positive score on Component 1 and a negative score on Component 2. This combination indicates companies that engage in extremely high-volume, promotional-focused email marketing but may not prioritize morning sending times. The negative loading on Component 2 suggests these companies might not strategically time their emails for the morning hours. These firms dominate in terms of sheer volume, indicating a strategy that prioritizes broad promotional outreach, likely at the expense of targeted or personalized engagement.

\end{itemize}

\begin{figure}[b]
    \centering
    \includegraphics[width=\linewidth]{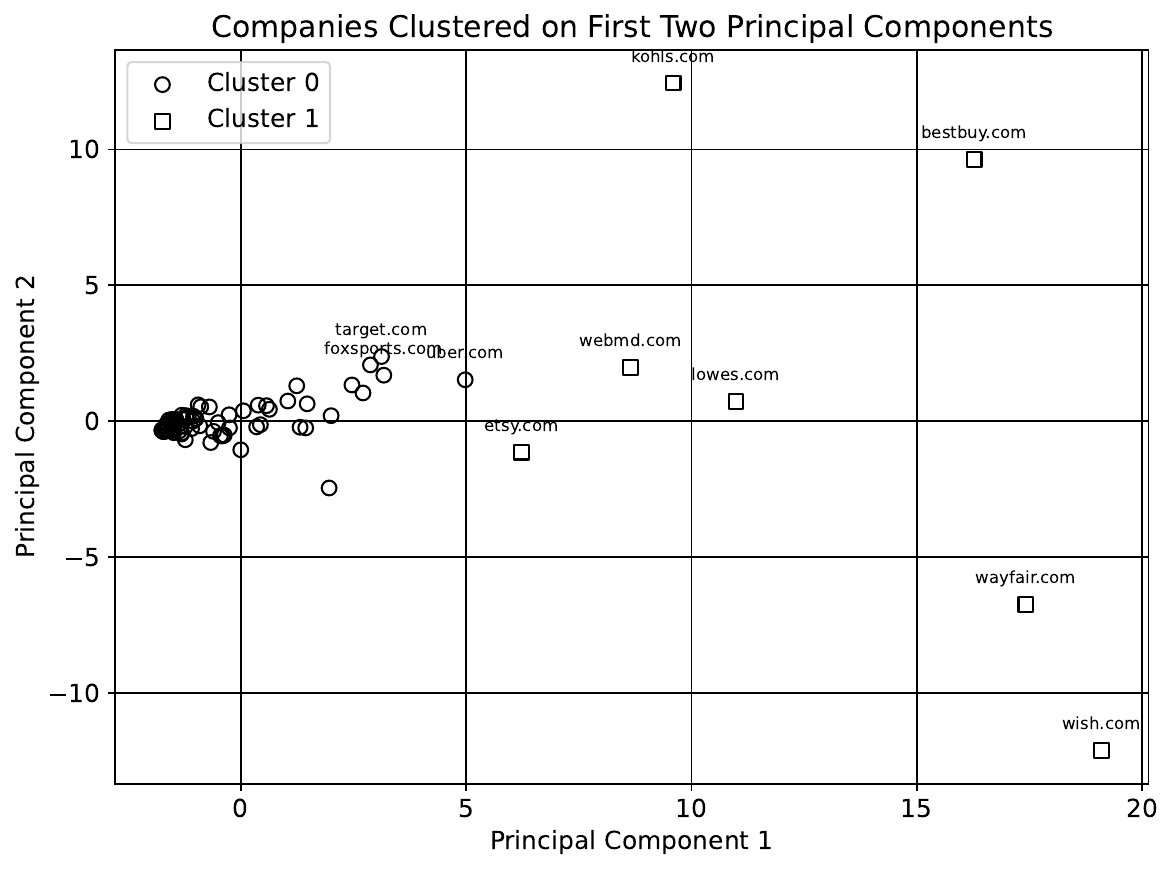}
    \caption{Cluster membership by top two principal components.}
    \label{fig:cluster_pca} 
\end{figure}

The clustering analysis, visually depicted in \cref{fig:cluster_pca}, reveals distinct segments of companies based on their email marketing strategies, ranging from high-volume promotional tactics to CRM-focused approaches. The PCA loadings provide a clear view of the factors driving these segmentations. From this analysis, there are two clear factor loadings that drive this clustering:

\begin{itemize}%
    \item \textbf{Component 1} captures overall email activity and promotional focus. Companies with high scores on this component send a high volume of emails, particularly promotional content, consistently across weekdays. This suggests a strategy that emphasizes frequent promotional outreach.
    
    \item \textbf{Component 2} represents email sending times, particularly in the morning hours. High scores on this component indicate companies that strategically send emails during morning hours (7 AM to 10 AM), possibly to increase open rates and customer engagement.
    
\end{itemize}

\begin{leftbar} \noindent
We see from this clustering that the dominant discriminator between these companies and their email sending patterns is the volume of specifically promotional emails that they are sending.
\end{leftbar}

\subsection{Cross-Sector Differences}

To explore potential differences in email sending behaviors across various sectors, we analyzed key metrics such as the total volume of emails sent, the types of emails (promotional, CRM, alert), and the timing of email sending (hourly and daily patterns). Since companies are moving to integrate their sales and marketing channels \cite{trenz2020disentangling}, and there are no uniformly accepted definitions that separate many of these companies \cite{lehrer2022omnichannel}, we rely on broad categorizations without making definitive claims as to their generalizability. This analysis is to further the argument, supported by our data, that not all companies send these unsolicited marketing emails in the same manner, and that certain sectors may be greater offenders than others. 

\textbf{Email Volume and Type by Sector.} A chi-squared test of independence was conducted to examine the relationship between email type and sector. The test revealed a statistically significant association between email type and sector (\( \chi^2(14) =2138.858, p < 0.0001 \)), indicating that the distribution of email types varies by sector, with results shown in \cref{tab:sector_email_volume}.

\begin{table}[ht]
\centering
\small
\caption{Sector email volume and type averages.}
\label{tab:sector_email_volume}
\begin{tabular}{lccccc}
\toprule
\textbf{Sector} & \textbf{Total} & \textbf{Promo} & \textbf{CRM} & \textbf{Alert} & \textbf{$n$} \\
\midrule
Omnichannel Retailers    & 139.56 & 133.22 & 5.89 & 0.44 & 9 \\
E-tailer                 & 123.23 & 104.69 & 17.31 & 0.38 & 13 \\
Financial Services       & 53.50 & 0.50 & 40.00 & 13.00 & 2 \\
Digital Services         & 23.69 & 7.89 & 15.17 & 0.60 & 35 \\
Online Marketplaces      & 26.75 & 20.06 & 6.63 & 0.00 & 16 \\
Communication Platforms  & 19.63 & 3.37 & 14.84 & 1.26 & 19 \\
Online Entertainment     & 15.20 & 5.10 & 9.10 & 1.00 & 10 \\
Brick \& Mortar          & 19.00 & 14.60 & 4.00 & 0.40 & 5 \\
\bottomrule
\end{tabular}
\end{table}

The significant chi-squared test suggests that sectors differ in the types of emails they send:

\begin{itemize}
    \item Retailers, both \textit{Omnichannel Retailers} and online only \textit{E-tailers}, predominantly send \textbf{promotional emails}. Examples of omnichannel companies include \textit{Kohls} and \textit{BestBuy}, while e-tailers include \textit{Wish.com} and \textit{Wayfair.com}. 
    
    \item The \textit{Financials} sector, while only 2 of the 21 companies actually sent emails, those two focus heavily on \textbf{CRM} (75.4\%) and \textbf{alert emails} (24.3\%) more than any other sector. This could reflect those companies using emails for non-marketing activities like account notifications. 
    
\end{itemize}

\textbf{Variability Among Individual Companies.} A one-way\linebreak ANOVA was conducted to compare the average emails sent per company by sector. The results indicated statistically significant difference (\( F(7, 101) = 3.5095, p = 0.002 \)), suggesting that the average email volume per company differs across sectors. Kruskal-Wallis H tests for both hourly and daily email sending patterns revealed no statistically significant differences between sectors. This indicates that sectors do not differ significantly in terms of when they send emails.

\begin{leftbar} \noindent
The significant differences in email volume and types between sectors implies that we may be able to direct future work towards those companies and sectors that are specifically sending promotional advertisements to consumers--retail and product-oriented businesses. While individual companies within the same sector may adopt diverse strategies in terms of email frequency and timing, there are some clear patterns that different industries utilize email communications to achieve their specific business objectives.
\end{leftbar}

\section{Discussion}
\label{sec-discussion}

\subsection{New Insights}

This study provides a comprehensive analysis of email marketing strategies employed by the most popular online websites and mobile applications. By collecting emails where each unique email address corresponds to a single company, we have ensured a precise lineage and isolation of source company to end user. This methodological approach, combined with a year-long email data collection period encompassing 150 apps and services, offers both breadth and depth to our dataset, distinguishing our work from prior studies. 
Our research contributes to the ongoing conversation about how companies utilize digital channels to market to and maintain relationships with their customers. The extensive dataset allows us to observe real-world practices over an extended period, providing insights into not only the frequency and content of communications but also the strategies underlying them.

\emph{The findings reveal no evidence of UUSS emails from unknown or unauthorized third-parties and a significant portion of consumer concerns regarding excessive promotional emails can be mitigated by opting out of marketing communications where possible.} The data also shows that the bulk of SOS emails originate from a small subset of companies, supporting the Pareto principle where approximately 80\% of the emails come from 20\% of the senders. This concentration suggests that \emph{aggressive email marketing is not a universal practice but is driven by specific companies}, often within certain sectors like retail.

\textbf{ASN Mapping Shows Reliance on Authorized Third-Parties.} 
A novel aspect of our study is the analysis of email provenance via the originating ASNs. By tracing the sender IP addresses to their corresponding ASNs, we uncover the reliance of companies on authorized third-party marketing and CRM services. Major intermediaries such as \textit{Salesforce}, \textit{Mailgun}, \textit{SendGrid}, and \textit{SparkPost} are identified as key players facilitating the bulk of email marketing traffic.

This approach sheds light on the ecosystem of email marketing infrastructure, highlighting how companies leverage specialized external platforms to execute their communication strategies. A complete picture, however, is elusive as many of the emails originate from public cloud providers such as \textit{AWS} and \textit{GCP}, so what third-party is operating its service hosted on these hyperscalers isn't able to easily be determined just from the IP address. It also raises questions about data privacy and the extent to which customer information is shared with third-party providers, a topic that warrants further investigation. 

\textbf{The ``What'' and ``When'' Matter.}
Our temporal analysis indicates that while the overall volume and timing of emails are relatively consistent across sectors, the content and purpose are tailored to industry-specific strategies. For instance, retailers predominantly sends promotional emails, aligning with their goal of stimulating immediate purchases. In contrast, financial service companies and communication platforms focus on CRM and alert emails, reflecting the importance of customer engagement and timely information dissemination.

The identification of distinct clusters through principal component analysis (PCA) and K-means clustering reveals that companies adopt varied email marketing tactics ranging from high-volume promotional outreach to personalized CRM communications. This diversity underscores the need for nuanced strategies that consider both the timing and content of emails to effectively engage customers.

\subsection{Legal and Ethical Considerations}

The findings raise important legal and ethical considerations regarding compliance with privacy policies and marketing preferences. Despite regulations like the CCPA, CAN-SPAM and GDPR, consumers continue to report receiving unsolicited emails. Our study suggests that opting out during registration can reduce the volume of such emails, but also highlights that some companies may not fully adhere to opt-out preferences. 

Aggressive marketing practices, especially those that aim to ``wear down'' consumers through persistent communications, can erode trust and infringe upon consumer rights. Prior research has demonstrated that excessive emailing can lead to customer fatigue and negative brand perception \cite{cao2019dynamic}. Companies must balance their marketing objectives with respect for consumer autonomy and privacy.

Policymakers and regulatory bodies may find these insights valuable in assessing the effectiveness of current regulations and identifying areas where enforcement or additional guidelines are necessary. Our study identifies that a vast majority of marketing emails are facilitated by third-party services such as \emph{Salesforce}, \emph{Mailgun}, and \emph{SendGrid}, with 89.35\% of emails being sent through the top 8 ASNs. \emph{This centralization indicates a potential focal point for improving regulatory oversight.} 
Policymakers should consider these patterns when evaluating regulatory frameworks, ensuring that opt-out mechanisms are straightforward, transparent, and honored in practice. Enhancing regulations to specifically reference and include the use of third-party marketing services and enforcing compliance can further improve consumer experiences and uphold consumer rights.

\subsection{Limitations and Future Work}

The richness of our dataset extends beyond metadata to include the full subject lines and body content of emails. This presents opportunities for future research into the specific language, persuasive techniques, and thematic elements employed in email marketing. To do this, future work can explore the content of emails in greater detail through Natural Language Understanding (NLU) techniques~\cite{allen1988natural,canonico2018comparison}. Additionally, most of the emails in the dataset contain images that could be rendered and then analyzed~\cite{zhang2023llavar,fei2024vitron}, yielding insights into specific persuasion techniques that are employed. Moreover, our dataset can support studies on the effectiveness of different email strategies by correlating communication patterns with customer engagement metrics, where available. Finally, investigating the role of cultural, regional, or demographic factors in email marketing practices also presents an avenue for further study.
Future work could also investigate whether smaller companies, which may face less regulatory scrutiny compared to popular online services and apps, are more likely to share or leak users' email addresses without consent.

While our study analyzes the correlation between spam reports and IP usage, identifying potential IP hopping behaviors, it does not yet evaluate whether such practices align with the companies’ stated privacy policies. Future work can focus on bridging this gap by cross-referencing email-sending behaviors with documented privacy commitments. This analysis would help determine if companies engaging in practices like IP hopping are acting inconsistently with their stated policies, providing deeper insights into accountability and compliance within the email ecosystem.

A key component of our framework is its low cost of ongoing maintenance and scalability. Once deployed, it can monitor any number of online services and apps passively. A key advantage of this is that we can detect any future changes in the inbound email messages. So for example, \emph{if a company in the monitored set has a breach that leaks customer email addresses and then those addresses are used for spam campaigns or sold to unauthorized third-parties, our system would be one of the first authoritative detection sources of such a breach}. 

\section{Conclusion}
\label{sec-conclusion}

This study offers a detailed examination of email inbox privacy and marketing practices among leading online services and apps, highlighting key patterns in communication strategies and the use of third-party services. Our findings contribute to a better understanding of how companies balance marketing objectives with consumer preferences and the implications for trust and engagement in the digital marketplace. By shedding light on these practices, we aim to inform both industry and policy discussions, promoting approaches that respect consumer rights while enabling effective email marketing communications. %

\bibliographystyle{ACM-Reference-Format}
\bibliography{references}

\appendix
\begin{appendix}
\section{LLM Classification Workflow}
\label{appendix-llm-classification}

The following code demonstrates the LLM classification workflow used to categorize emails as promotional, CRM, or alert.

\begin{lstlisting}[style=python, caption={LLM classifier prompt and code.}, basicstyle=\scriptsize\ttfamily]
tagging_prompt = ChatPromptTemplate.from_template(
    """
You are an email classifier system. Your task is to classify the following email message as either promotional, CRM, or alert. Respond only with the JSON formatted classification based on the provided schema below, containing only the classification results and not the schema itself. Do NOT output the schema or any definitions-only provide the classification response.

Here is the JSON schema of the response you must strictly follow:

{schema}

Here are the classification definitions (these are for your reference only, do not include them in the output):

promotional: Emails offering price discounts intended to induce short-term actions such as purchasing. These are overt selling attempts with a call to action.

CRM: Emails that engage, inform, or enhance the company's relationship with the customer. These emails are not overtly sales-focused but aim for longer-term goals, like brand building or engagement.

alert: Emails containing various notifications, updates, or alerts based on consumer preferences. These are informational and do not directly attempt to sell.

Classify the following email strictly based on these definitions:

Email:
{input}
"""
)

# Create the structured output format class
class Classification(BaseModel):
    sentiment: str = Field(..., enum=["promotional", "CRM", "alert"])
    confidence: int = Field(
        ...,
        description="describes how confident the classification is, the higher the number the more confident in the classification",
        enum=[1, 2, 3, 4, 5],
    )
    rationale: str = Field(description="The primary driver of the classification based on the definitions")
\end{lstlisting}

{\onecolumn 
\section{List of Received Email Domains}
\label{appendix-domains}

\cref{table:all-domains} shows the details and distribution of emails received from 109 different root domains.

\small
\begin{longtable}{p{3cm}p{4cm}p{1.5cm}p{1.5cm}p{1.5cm}p{1.5cm}p{1.5cm}}
\caption{Distribution of emails received from 109 different root domains.}
\label{table:all-domains} \\
\toprule
Root Domain & Sector & Cluster & Total Emails & Promotional & CRM & Alert \\
\midrule
\endfirsthead
\toprule
Root Domain & Sector & Cluster & Total Emails & Promotional & CRM & Alert \\
\midrule
\endhead
\midrule
\multicolumn{7}{r}{Continued on next page} \\
\midrule
\endfoot
\bottomrule
\endlastfoot
adobe.com & Digital Services & 0 & 37 & 24 & 13 & 0 \\
airtable.com & Digital Services & 0 & 2 & 0 & 2 & 0 \\
aliexpress.com & E-tailer & 0 & 37 & 33 & 3 & 1 \\
apartmentfinder.com & Online Marketplace & 0 & 1 & 0 & 1 & 0 \\
apartments.com & Online Marketplace & 0 & 8 & 1 & 7 & 0 \\
autozone.com & Brick and Mortar & 0 & 13 & 7 & 5 & 1 \\
bandcamp.com & Digital Services & 0 & 4 & 1 & 3 & 0 \\
bestbuy.com & Omnichannel & 1 & 391 & 376 & 12 & 3 \\
bloomberg.com & Financials & 0 & 8 & 0 & 5 & 3 \\
bloombergbusiness.com & Financials & 0 & 99 & 1 & 75 & 23 \\
booking.com & Online Marketplace & 0 & 2 & 0 & 2 & 0 \\
canva.com & Digital Services & 0 & 3 & 1 & 2 & 0 \\
capcut.com & Communication Platforms & 0 & 71 & 19 & 49 & 1 \\
cars.com & E-tailer & 0 & 15 & 6 & 9 & 0 \\
cnet.com & Digital Services & 0 & 1 & 0 & 1 & 0 \\
codepen.io & Digital Services & 0 & 49 & 1 & 47 & 1 \\
codewars.com & Digital Services & 0 & 3 & 0 & 3 & 0 \\
coursehero.com & Digital Services & 0 & 22 & 7 & 15 & 0 \\
discogs.com & Communication Platforms & 0 & 8 & 0 & 7 & 1 \\
discord.com & Communication Platforms & 0 & 3 & 0 & 3 & 0 \\
doordash.com & Online Marketplace & 0 & 3 & 3 & 0 & 0 \\
dunkinrewards.com & Brick and Mortar & 0 & 70 & 61 & 8 & 1 \\
ebay.com & E-tailer & 0 & 70 & 57 & 13 & 0 \\
etsy.com & E-tailer & 1 & 196 & 115 & 81 & 0 \\
expedia.com & Digital Services & 0 & 28 & 24 & 3 & 1 \\
flickr.com & Communication Platforms & 0 & 44 & 13 & 31 & 0 \\
foodnetwork.com & Online Entertainment & 0 & 1 & 0 & 1 & 0 \\
foxsports.com & Online Entertainment & 0 & 111 & 39 & 63 & 9 \\
fragrancenet.com & E-tailer & 0 & 50 & 41 & 9 & 0 \\
freecodecamp.org & Digital Services & 0 & 1 & 0 & 1 & 0 \\
geeksforgeeks.org & Digital Services & 0 & 12 & 7 & 5 & 0 \\
github.com & Digital Services & 0 & 2 & 0 & 1 & 1 \\
google.com & Communication Platforms & 0 & 3 & 0 & 2 & 1 \\
hackerrankmail.com & Digital Services & 0 & 4 & 0 & 4 & 0 \\
homedepot.com & Omnichannel & 0 & 1 & 0 & 1 & 0 \\
hotels.com & Online Marketplace & 0 & 89 & 72 & 17 & 0 \\
imdb.com & Online Entertainment & 0 & 2 & 0 & 2 & 0 \\
instagram.com & Communication Platforms & 0 & 14 & 0 & 13 & 1 \\
kayak.com & Digital Services & 0 & 3 & 0 & 3 & 0 \\
khanacademy.org & Digital Services & 0 & 49 & 21 & 28 & 0 \\
kohls.com & Omnichannel & 1 & 335 & 330 & 5 & 0 \\
leetcode.com & Digital Services & 0 & 8 & 5 & 3 & 0 \\
linkedin.com & Communication Platforms & 0 & 31 & 1 & 30 & 0 \\
lowes.com & Omnichannel & 1 & 411 & 404 & 6 & 1 \\
lyftmail.com & Online Marketplace & 0 & 7 & 1 & 6 & 0 \\
nationalgeographic.com & Online Entertainment & 0 & 1 & 0 & 1 & 0 \\
newegg.com & E-tailer & 0 & 1 & 1 & 0 & 0 \\
nike.com & Omnichannel & 0 & 1 & 0 & 1 & 0 \\
openai.com & Digital Services & 0 & 3 & 0 & 3 & 0 \\
opentable.com & Digital Services & 0 & 2 & 0 & 2 & 0 \\
oracle-mail.com & Digital Services & 0 & 5 & 1 & 4 & 0 \\
oracle.com & Digital Services & 0 & 1 & 0 & 1 & 0 \\
orbitz.com & Online Marketplace & 0 & 15 & 14 & 1 & 0 \\
pandora.com & Communication Platforms & 0 & 36 & 17 & 19 & 0 \\
pinterest.com & Communication Platforms & 0 & 33 & 1 & 32 & 0 \\
planetfitness.com & Brick and Mortar & 0 & 1 & 0 & 1 & 0 \\
plex.tv & Online Entertainment & 0 & 4 & 0 & 4 & 0 \\
pons.com & Digital Services & 0 & 1 & 0 & 1 & 0 \\
pornhub.com & Online Entertainment & 0 & 3 & 0 & 3 & 0 \\
quizlet.com & Digital Services & 0 & 5 & 0 & 5 & 0 \\
quora.com & Digital Services & 0 & 67 & 0 & 63 & 3 \\
realtor.com & Online Marketplace & 0 & 7 & 2 & 5 & 0 \\
reddit.com & Communication Platforms & 0 & 1 & 0 & 1 & 0 \\
redditmail.com & Communication Platforms & 0 & 86 & 0 & 65 & 20 \\
reuters.com & Communication Platforms & 0 & 2 & 0 & 2 & 0 \\
ring.com & Omnichannel & 0 & 5 & 1 & 4 & 0 \\
roblox.com & Online Entertainment & 0 & 1 & 0 & 1 & 0 \\
roku.com & Communication Platforms & 0 & 3 & 0 & 3 & 0 \\
scrimba.com & Digital Services & 0 & 17 & 12 & 4 & 1 \\
shazam.com & Communication Platforms & 0 & 1 & 0 & 1 & 0 \\
shein.com & E-tailer & 0 & 106 & 93 & 12 & 1 \\
sheinnotice.com & E-tailer & 0 & 1 & 0 & 1 & 0 \\
shopify.com & Online Marketplace & 0 & 1 & 0 & 1 & 0 \\
shutterstock.com & Digital Services & 0 & 52 & 20 & 31 & 1 \\
skillcrush.com & Online Entertainment & 0 & 27 & 12 & 15 & 0 \\
snapchat.com & Communication Platforms & 0 & 5 & 0 & 5 & 0 \\
soundcloud.com & Communication Platforms & 0 & 11 & 3 & 8 & 0 \\
southwest.com & Brick and Mortar & 0 & 1 & 0 & 1 & 0 \\
springernature.com & Omnichannel & 0 & 2 & 0 & 2 & 0 \\
statista.com & Digital Services & 0 & 1 & 0 & 1 & 0 \\
steampowered.com & E-tailer & 0 & 1 & 0 & 1 & 0 \\
target.com & Omnichannel & 0 & 109 & 87 & 22 & 0 \\
temu.com & E-tailer & 0 & 3 & 0 & 3 & 0 \\
temuemail.com & E-tailer & 0 & 61 & 40 & 20 & 1 \\
thumbtack.com & Online Marketplace & 0 & 11 & 1 & 10 & 0 \\
tiktok.com & Communication Platforms & 0 & 1 & 0 & 1 & 0 \\
today.com & Digital Services & 0 & 116 & 112 & 3 & 1 \\
topcoder.com & Digital Services & 0 & 1 & 0 & 1 & 0 \\
tracfone.com & Brick and Mortar & 0 & 10 & 5 & 5 & 0 \\
trip.com & Online Marketplace & 0 & 2 & 0 & 2 & 0 \\
tripadvisor.com & Online Marketplace & 0 & 74 & 32 & 42 & 0 \\
trivago.com & Online Marketplace & 0 & 32 & 26 & 6 & 0 \\
turo.com & Online Marketplace & 0 & 1 & 0 & 1 & 0 \\
tutorialsdojo.com & Digital Services & 0 & 21 & 15 & 5 & 1 \\
tutorialspoint.com & Digital Services & 0 & 23 & 16 & 7 & 0 \\
twitch.tv & Communication Platforms & 0 & 6 & 0 & 6 & 0 \\
uber.com & Online Marketplace & 0 & 174 & 169 & 4 & 0 \\
udacity.com & Digital Services & 0 & 1 & 0 & 1 & 0 \\
usnews.com & Digital Services & 0 & 2 & 0 & 2 & 0 \\
vimeo.com & Communication Platforms & 0 & 14 & 10 & 4 & 0 \\
walmart.com & Omnichannel & 0 & 1 & 1 & 0 & 0 \\
wayfair.com & E-tailer & 1 & 559 & 520 & 27 & 2 \\
waze.com & Digital Services & 0 & 2 & 0 & 2 & 0 \\
webmd.com & Digital Services & 1 & 270 & 9 & 250 & 11 \\
wish.com & E-tailer & 1 & 502 & 455 & 46 & 0 \\
xhamster.com & Online Entertainment & 0 & 1 & 0 & 1 & 0 \\
xvideos.com & Online Entertainment & 0 & 1 & 0 & 0 & 1 \\
yelp.com & Digital Services & 0 & 11 & 0 & 11 & 0 \\
zillow.com & Online Marketplace & 0 & 1 & 0 & 1 & 0 \\
\end{longtable}
}

\end{appendix}

\end{document}